\documentclass{iau}
\usepackage{graphicx,natbib}
 
\newcommand{\apj}{ApJ}           
\newcommand{\mnras}{MNRAS}       

\title
[Stellar masses from micro-lensed quasars]
{Stellar masses calibrated \\ with micro-lensed quasars}

\author
[Schechter, Blackburne, Pooley \& Wambsganss]
{Paul L. Schechter$^1$, Jeffrey A. Blackburne$^2$,
David Pooley$^3$ and Joachim Wambsganss$^4$}

\affiliation{$^1$MIT Kavli Institute Cambridge, MA 02139, USA \\ 
email: {\tt schech@mit.edu}\\
$^2$ Aret\'e Associates, Northridge, CA 91324\\
$^3$ Sam Houston State University, Huntsville, TX 77341\\
$^4$ Zentrum f\"ur Astronomie der Universit\"at Heidelberg, Germany
}

\pubyear{2014}
\volume{311}  
\jname{Galaxy Masses as Constraints of Formation Models}
\editors{M. Cappellari \& S. Courteau, eds.}
\begin{document}

\maketitle

\begin{abstract}
We measure the stellar mass surface densities of early type galaxies
by observing the micro-lensing of macro-lensed quasars caused by
individual stars, including stellar remnants, brown dwarfs and red
dwarfs too faint to produce photometric or spectroscopic signatures.
Our method measures the {\it graininess} of the gravitational
potential, in contrast to methods that decompose a smooth total
gravitational potential into two smooth components, one stellar and
one dark.  We find the median likelihood value for the calibration
factor $\cal F$ by which Salpeter stellar masses (with a low mass cutoff of
$0.1 M_\odot$) must be multiplied is 1.23, with a one sigma confidence
range of $0.77 < {\cal F} < 2.10$.

\end{abstract}

\firstsection 

\section{Introduction}

Stellar masses for early type galaxies are almost always determined by
one of two methods: either they are estimated from spectra (and sometimes
only broad band colors) or they are deduced by subtracting the
contribution of an assumed dark matter component from a combined mass
inferred from kinematic (and sometimes macro-lensing) measurements.
Multiple examples of both methods can be found in the present volume.
Both methods have shortcomings.

Here we use a third method: determining the stellar mass surface
density of an early type galaxy from brightness fluctuations of the
four images of a background quasar that is both multiply-imaged
(``macro-lensed'') by the galaxy and ``micro-lensed'' by the
individual stars in that galaxy \citep{Schechter2004, Kochanek2004}.
This method, in
contrast to spectral methods, is sensitive to stellar mass near and
below the hydrogen burning limit, as well as to the mass in stellar
remnants.  And where dark matter subtraction methods make asumptions
about the dark matter profiles, the gravitational micro-lensing
technique makes only an assumption about the combined gravitational
potential, one that has been subjected to extensive observational
verification.

Ideally one would observe a single system long enough to see a great
many fluctuations and infer an accurate stellar surface density.  But
the timescale for micro-lensing variations is of order ten years for a
lens at redshift $z \sim 0.5$ \citep{Mosquera2011}, and observations
of even four quasar images at a single epoch give only broad
constraints on the mass surface density.  So instead we observe ten
systems at a single epoch and combine results.

\section{Calibrating the stellar mass fundamental plane}

Our approach is to use the stellar mass fundamental plane \citep{Hyde2009}
 to predict the stellar surface mass densities at the
positions of our quasar images, which cause differences between the
observed quasar fluxes and those predicted by a macro-model.  We
then adjust the stellar masses by a multiplicative constant, $\cal F$.
The constant is varied and the likelihood of the observed quasar
X-ray fluxes is computed for each value of $\cal F$.

\begin{figure}[t]
 \centerline{
 \scalebox{0.55}{
 \includegraphics[angle=270]{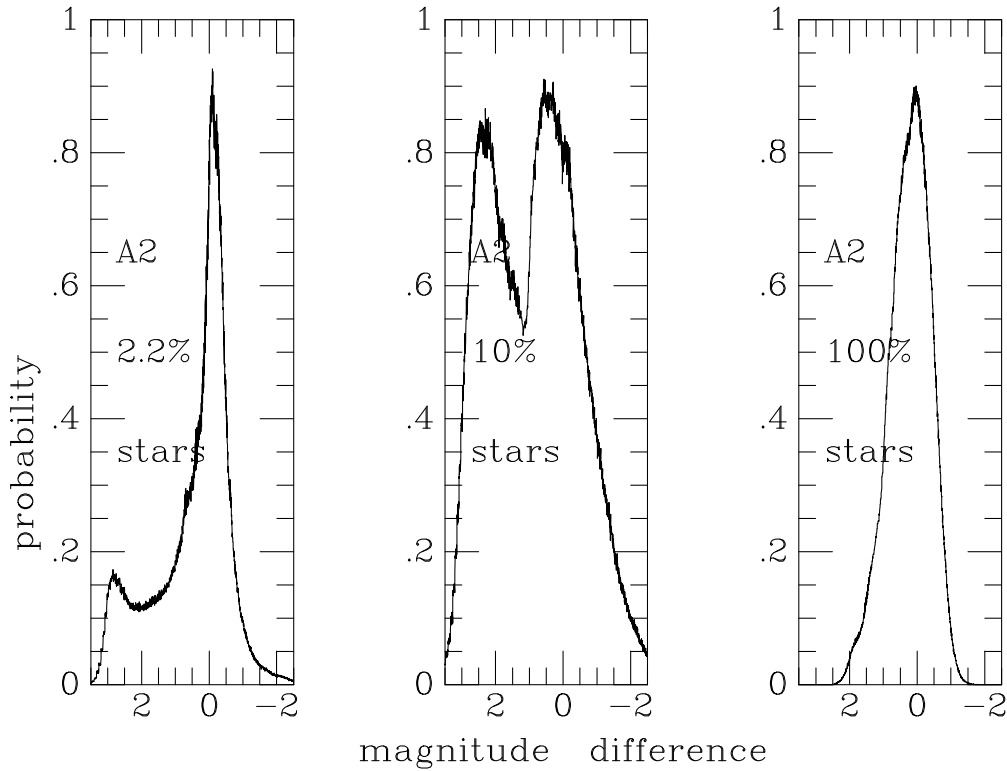}}
 }
  \caption{
Probability distribution for the ratio
of observed to macro-model flux (expressed as a magnitude difference) 
for the $A2$  image of the quadruple lens PG 1115+080 for three
different stellar mass fractions.
The different shapes of the distributions
permit determination of  the stellar mass fraction.
  }\label{fig:pg1115}
\end{figure}

Our stellar mass fundamental plane was constructed using data from
\cite{Auger2010} and 
\cite{Sonnenfeld2013}, who used a Salpeter
initial mass function (IMF) with a low mass cutoff of $0.1M_\odot$
to compute stellar masses for lensing galaxies at $z \sim 0.2$ and
$z \sim 0.5$ respectively.  The median likelihood value of
the factor by which these masses must be multiplied is ${\cal F} = 1.23$
with a 68\% confidence interval of $0.77 < {\cal F} < 2.10.$  The
range is the result of small sample size.

\begin{figure}[t]
 \centerline{
 \scalebox{0.55}{
 \includegraphics[angle=270]{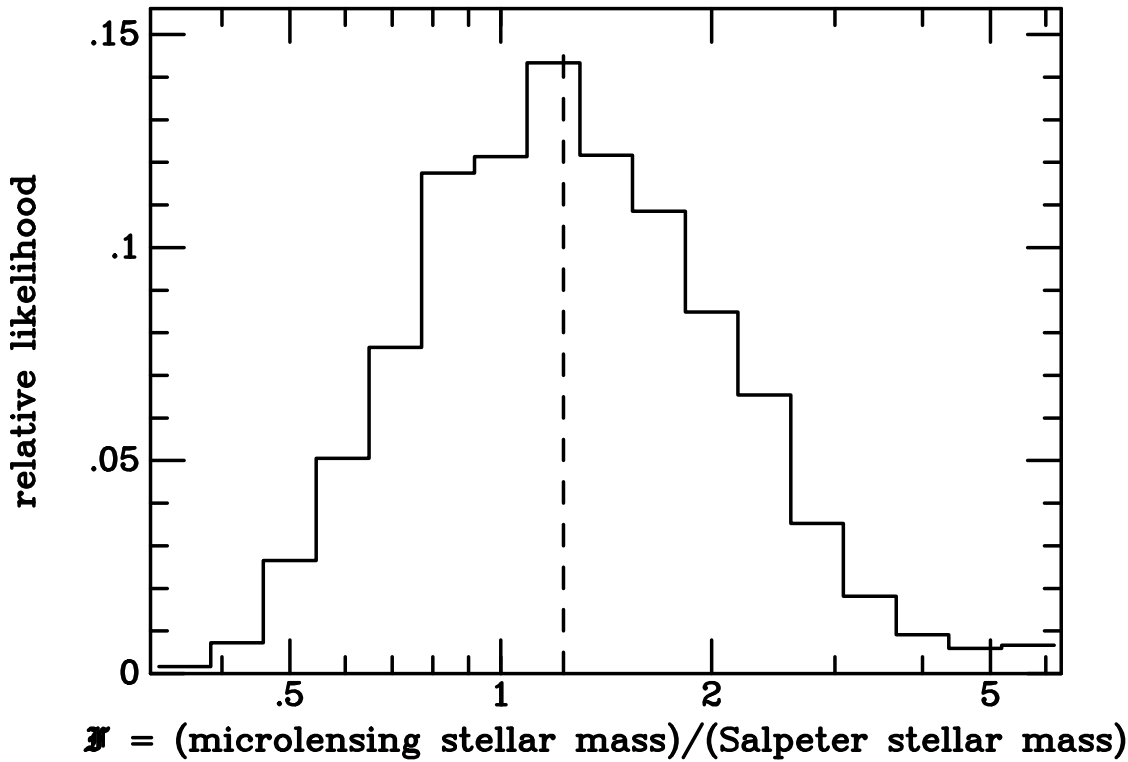}}
 }
  \caption{
  The likelihood histogram for the calibration factor
  $\cal F$ by which the stellar mass fundamental
  plane computed using a Salpeter IMF must be multiplied
  to produce the observed X-ray flux ratio anomalies.
  }\label{fig:likeli}
\end{figure}

\section{Salient features of the analysis}

\noindent
$\bullet$ We take the quasar to be point-like 
 at X-ray wavelengths
(relative to the
Einstein rings of the micro-lensing stars) and
use Chandra fluxes taken from \cite{Pooley2007}.
\cite{Blackburne2011} have shown that the optical and near-IR
emitting regions are comparable in projected size to the stellar
Einstein rings, rendering optical wavelengths less well suited to the
present analysis.

$\bullet$ We use a fundamental plane in which a ``proxy'' velocity
dispersion computed from the radius of the Einstein ring of the
macro-model is substituted for stellar dispersions.  We find that for
the \cite{Auger2010} sample this gives
considerably less scatter than using proper stellar velocity
dispersions in constructing the fundamental plane.

$\bullet$ We model the lenses as singular isothermal ellipsoids
(ellipticity and orientation identical to that of the 
stellar light) with an external shear.  

$\bullet$ We have analyzed the sensistivity of our result
to a number of different possible sources of systematic error.
We find them to be small compared to the statistical confidence
interval.  In particular we find that milli-lensing by dark
matter substructure has very little effect on our result.

$\bullet$ Our result is most sensitive to the measured de Vaucouleurs
radii of the lensing galaxies.  The published values fall into two
groups, based on the software used, which deviate systematically from
each other; we adopt the geometric mean.

$\bullet$ An alternative analysis that takes surface mass density
and effective radius to be functions of the proxy velocity dispersion --
a ``fundamental line'' -- circumvents the systematic
uncertainties in the measured effective radii and yields a calibration factor
$\cal F$ that differs by 7\%.


$\bullet$ The X-ray fluxes for the images of RX J0911+0551 pull our
calibration factor ${\cal F}$ to higher values.  The macro-models
predict that image $b$, a minimum of the light travel time, should be
more than five times brighter than image $d$, also a minimum, but is
observed to be only 25\% brighter.


$\bullet$ For a galaxy
with a proxy velocity dispersion of 266 km/s and an effective radius
of 6.17 kpc, average for the Auger et al. (2010) sample,
the calibrated fractional stellar surface mass density
peaks at a value of 1.10 at $0.074 r_e$, as shown in Figure 3.
At $1.5 r_e$, the typical Einstein radius for our lens systems,
the stellar mass fraction is 0.25.

\begin{figure}[t]
 \centerline{
 \scalebox{0.45}{
 \includegraphics[angle=270]{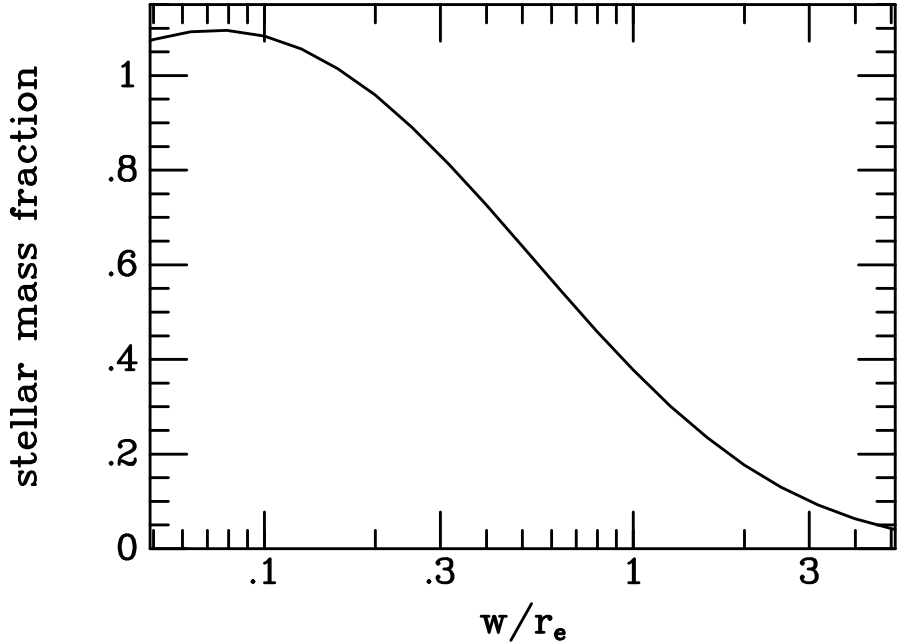}}
 }
\caption{The stellar surface mass density as a fraction of the
total for the typical lensing galaxy, obtained by applying
the calibration factor $\cal{F}$ to the stellar mass
fundamental plane derived from the SLACS + SL2S samples.
The peak at $w/r_e = 0.074$ (where $w$ is the circularized
radius)  occurs where the slope of the  de Vaucouleurs
profile is equal to that of the singular isothermal.
}
\label{fig:darkfrac}
\end{figure}

\medskip
The work summarized here is the culmination of an effort begun by
\cite{Schechter2004} and described in
the proceedings of IAU Symposium 220.  A more thorough description of
the present analysis can be found in \cite{Schechter2014}

\begin{acknowledgments}
The authors gratefully acknowledge the very substantial
contributions of Saul A. Rappaport.
This work was supported in part by the US National 
Science Foundation under grants AST02-06010 and AST06-07601
and by NASA under Chandra grant G07-8099. 
The first author thanks the members
of the organizing committee for their good efforts.
\end{acknowledgments}

\end{document}